\documentclass[preprint,12pt,3p]{elsarticle}
\journal{Ultramicroscopy}

\usepackage{lineno,hyperref}
\usepackage{amsfonts}
\usepackage{amsmath}
\usepackage{gensymb}
\usepackage{amssymb}
\usepackage{siunitx} % use this package module for SI units
\newcommand{\angstrom}{\textup{\AA}}
\usepackage[normalem]{ulem}
\usepackage{caption}
\usepackage{subcaption}
\usepackage{dcolumn}
\usepackage{bm}
\usepackage{color}
\usepackage{hyperref}
\usepackage{bookmark}
\usepackage{empheq}
\usepackage{booktabs}
\usepackage{multirow}
\usepackage{pdfpages}
\usepackage{graphicx}
\usepackage{tabu}
\usepackage[version=4]{mhchem}
\usepackage{soul,xcolor}
\usepackage[utf8]{inputenc}
\usepackage{epigraph}
\usepackage{physics}
\usepackage{textcomp}
\usepackage{tikz}
\usepackage{notoccite}
\usetikzlibrary{shapes.geometric, arrows}
\usepackage[T1]{fontenc}
\usepackage{booktabs}

\begin{document}

\begin{frontmatter}

\title{Fast generation of calculated ADF-EDX scattering cross-sections under channelling conditions}

%% Group authors per affiliation:
\author[EMAT,Nano,Oxford]{Zezhong Zhang}\corref{cor1}
\ead{zezhong.zhang@uantwerpen.be}
\author[EMAT,Nano]{Ivan Lobato}
\author[EMAT,Nano]{Annick De Backer}
% \ead{annick.debacker@uantwerpen.be}
% \ead{Ivan Lobato@uantwerpen.be}
\author[EMAT,Nano]{Sandra Van Aert}\corref{cor2}
\ead{sandra.vanaert@uantwerpen.be}
\author[Oxford]{Peter Nellist}\corref{cor3}
\ead{peter.nellist@materials.ox.ac.uk}

\address[EMAT]{ Electron Microscopy for Materials Research (EMAT), University of Antwerp, Groenenborgerlaan 171, 2020 Antwerp, Belgium}
\address[Nano]{NANOlab Center of Excellence, University of Antwerp, Groenenborgerlaan 171, 2020 Antwerp, Belgium }
\address[Oxford]{Department of Materials, University of Oxford, 16 Parks Road, Oxford OX1 3PH, United Kingdom}
\cortext[cor1]{Corresponding author}

\begin{abstract}
Advanced materials often consist of multiple elements which are arranged in a complicated structure. Quantitative scanning transmission electron microscopy is useful to determine the composition and thickness of nanostructures at the atomic scale. However, significant difficulties remain to quantify mixed columns by comparing the resulting atomic resolution images and spectroscopy data with multislice simulations where dynamic scattering needs to be taken into account. The combination of the computationally intensive nature of these simulations and the enormous amount of possible mixed column configurations for a given composition indeed severely hamper the quantification process. To overcome these challenges, we here report the development of an incoherent non-linear method for the fast prediction of ADF-EDX scattering cross-sections of mixed columns under channelling conditions. We first explain the origin of the ADF and EDX incoherence from scattering physics suggesting a linear dependence between those two signals in the case of a high-angle ADF detector. Taking EDX as a perfect incoherent reference mode, we quantitatively examine the ADF longitudinal incoherence under different microscope conditions using multislice simulations. Based on incoherent imaging, the atomic lensing model previously developed for ADF is now expanded to EDX, which yields ADF-EDX scattering cross-section predictions in good agreement with multislice simulations for mixed columns in a core-shell nanoparticle and a high entropy alloy. The fast and accurate prediction of ADF-EDX scattering cross-sections opens up new opportunities to explore the wide range of ordering possibilities of heterogeneous materials with multiple elements.
\end{abstract}


\end{frontmatter}

% \linenumbers

\section{Introduction}
Despite their small size, nanostructured materials can display extraordinarily complex atomic structures associated with chemical inhomogeneities. Since their properties are fundamentally determined by the exact atomic arrangement, a quantitative structural characterisation in 3D is essential to get insight into the structural-properties relationship and hence the development of next-generation nanostructured materials. A popular characterisation technique is annular dark field scanning transmission electron microscopy (ADF-STEM) because of its sub-angstrom resolution in combination with its sensitivity to both the sample thickness and atomic number. To retrieve the 3D atomic structure, one can tilt the sample to different viewing directions and perform electron tomography. State-of-the-art ADF-STEM tomography has reached atomic resolution \cite{Goris2015,Yang2017b}. In addition, from a single ADF-STEM image, it also has been demonstrated that one can determine the atomic column positions and count the number of atoms with high precision and accuracy for homogeneous materials \cite{Lebeau2010,DeBacker2013}. In combination with prior knowledge about the crystal periodicity along the electron beam direction, atom counts can be translated into an initial atomic model, which can be further optimised using an energy minimisation algorithm to obtain a low energy state of the nanostructure \cite{arslan20213}. A quantitative comparison study showed an excellent agreement between atomic resolution electron tomography and atom counting reconstructions \cite{DeBacker2017}. This method is dose-efficient since it only requires a single viewing direction. Therefore, it is suitable for 3D characterisation of beam sensitive materials and for the investigation of particle dynamics at the atomic scale during in-situ experiments \cite{Bals2012, altantzis2018three, Liu2021}. 

To count the number of atoms from ADF-STEM images, we measure the so-called scattering cross-section (SCS), corresponding to the total intensities of electrons scattered by a single atomic column within the angular range of the ADF detector \cite{VanAert2009,E2013}. This quantity outperforms peak intensities because of its monotonic increase against the sample thickness and robustness against various probe conditions (including defocus, source coherence and aberrations) \cite{E2013}. In practice, scattering cross-sections are measured by integrating the STEM signal over the Voronoi cell for each atomic column \cite{Jones2014} or by estimating the volume under a Gaussian peak that models an atomic column shape \cite{DeBacker2016}. If the experimental images are normalised against the incident beam, the resulting scattering cross-sections can be quantitatively compared with simulated libraries obtained under the same experimental conditions, enabling us to count the number of atoms in the viewing direction for homogeneous materials. Alternatively, Van Aert et al. \cite{VanAert2011a,VanAert2013a,DeBacker2013a} proposed a statistics-based method which decomposes the distribution of scattering cross-sections into overlapping normal distributions each corresponding to a specific number of atoms. One may further combine the simulation and statistics-based method for a more reliable structural quantification \cite{VanAert2013a,Dewael2017}. For heterogeneous materials, the solution is often constrained in previous studies \cite{VanAert2009,Martinez2014a} by assuming a constant thickness and a linear dependence of the scattering cross-sections on the chemical composition. However, this is only an approximation since the scattering cross-sections depend on the location and the ordering of atoms in the column \cite{voyles2002atomic,hwang2013three,ishikawa2014three,Akamine2015,MacArthur2017,VandenBos2019a}. Based on the channelling theory of incoherent imaging, van den Bos et al. \cite{VanDenBos2016a,VandenBos2019a} developed the so-called atomic lensing model to take the ordering of multiple elements into account. This model predicts the ADF scattering cross-section of a mixed column from the libraries of pure elements. When including a priori knowledge about the sample, this was successfully applied to count the number of atoms for an Au@Ag core-shell nanorod \cite{VandenBos2019a, VanDenBos2016a}. To overcome the need for a priori knowledge and to unscramble binary systems with mixed elements which are close in atomic number (Pt-Au for example), it is difficult to rely on ADF-STEM images alone.

Energy dispersive X-ray (EDX) spectroscopy and electron energy loss spectroscopy (EELS) can fingerprint different elements. With modern instrumentation, the acquisition of EDX and EELS spectrum imaging datasets at atomic resolution is now becoming more routinely possible. The synchronisation of the signals between the probe scanning system and different detectors allows simultaneous acquisition of ADF-EDX-EELS hence maximising the transfer of structural and chemical information \cite{longo2013fast,Thersleff2020}. In addition, fast-scan multi-frame imaging techniques can mitigate scan noise (both linear and non-linear), reduce the sample damage, and improve the signal-to-noise ratio \cite{Jones2018,Wang2018a}. The fast-evolving detector design also leads to an ever-changing detector geometry and efficiency \cite{zaluzec2022quantitative}, which needs to be accounted for quantitatively when calibrating EDX signals to the absolute scale \cite{Xu2016a,Kraxner2017,Xu2018}. To overcome the difficulties in the characterisation of the EDX detectors, we can incorporate the experimentally measured EDX partial cross-section, which is called a \textit{partial} scattering cross section since it includes the microscope-dependent factors during normalisation \cite{Varambhia2018}. \par 

Even though atomic resolution spectroscopy has gradually improved from the experimental side and inelastic scattering calculations within the multislice framework are well-established (see review \cite{Dwyer2013} and references therein), difficulties for quantification persist. Dynamical scattering results in a non-linear response for elements at different depths, thus significantly complicating composition quantification. If we want to quantify spectroscopy data alongside ADF using similar quantification routines, we need to include the effects of channelling in the spectroscopy simulations. Though both high-angle ADF and EDX are known to be highly localised and incoherent, it is unclear whether they follow the same channelling behaviour. In fact, since the EDX signal is fully incoherent, the EDX-ADF comparison allows an investigation of the degree of ADF longitudinal incoherence \cite{nellist2000principles}, which is largely unexplored. In addition, the number of possible configurations grows exponentially with the number of different types of elements and thickness of the sample, hence quickly exceeding the computation time of multislice calculations. Therefore, MacArthur et al. \cite{MacArthur2017,MacArthur2021} suggested tilting the sample by ~2-3$\degree$ to reduce the effect of channelling to perform EDX quantification, which is at the cost of resolution. To have both the atomic resolution and computational feasibility in the presence of channelling, the applicability of the atomic lensing model to efficiently predict EDX scattering cross-sections of mixed columns will be investigated. This model has previously been developed to predict ADF scattering cross-sections of mixed columns \cite{VandenBos2019a, VanDenBos2016a}. Since its origin is based on longitudinal incoherent imaging, it is expected that this method will be applicable for fast EDX predictions.\par

The present paper aims to address the following key questions related to ADF-EDX quantification under channelling conditions: (a) Do EDX and ADF scattering cross sections have the same thickness scaling behaviour due to channelling? (b) How does the longitudinal incoherence of ADF compare to EDX? (c) How can the atomic lensing model be used to  predict EDX scattering cross-sections for mixed columns? In section~\ref{sec:theory}, we will discuss the origin of the incoherence for ADF and EDX signals in the multislice framework. In section~\ref{sec:dependence}, we will examine the longitudinal incoherence of ADF signals by simulating the ADF-EDX scattering cross-sections under different microscope conditions. In section~\ref{sec:ALM}, we will expand the atomic lensing model to spectroscopy enabling a fast prediction of EDX scattering cross-sections of mixed columns.

\section{Electron scattering theory for ADF and EDX within the multislice framework} 
\label{sec:theory}
By dividing materials into slices, the multislice algorithm describes multiple scattering as a repetition of single scattering within each slice and free propagation between slices. In this section, we will briefly outline the equations for ADF and EDX signals to understand their relationship, while readers are referred to Kirkland's book on the full topics of multislice \cite{Kirkland2010} and the review by Dwyer on the inelastic scattering \cite{Dwyer2013}. 

The relativistically-corrected Schrödinger equation for a fast electron travelling in the forward direction z \cite{howie1968approximations} can be written as: 
\begin{equation}
    \frac{\partial \psi(r, R, z)}{\partial z} = [\frac{i\lambda}{4\pi}(\nabla^{2}_{r})+ i \sigma V(r,z)] \psi(r, R, z),
\label{eq:wave_evo}
\end{equation}
where $\psi(r, R, z)$ is the electron wave at thickness $z$, probe position $R$ and real space 2-D coordinate vector $r = (x,y)$. The impact parameter is $\sigma=m e \lambda  /2\pi\hbar^2$, $V(r,z)$ is the electrostatic potential at depth $z$, $e$ is the electron charge, $m$ and $\lambda$ are the relativistically corrected electron mass and wavelength, respectively. Once the electron wave reaches the exit surface, it propagates to the detector plane in the far-field. The intensity scattered within the inner and outer collection angle of the ADF detector will be collected: \begin{equation}
    I_{ADF}(R) = \int D(k)|\psi(k, R, z)|^2 dk^2,
\label{eq:adf}
\end{equation}
where $\psi(k, R, z)$ is the Fourier transform of $\psi(r, R, z)$, $D(k)$ is the ADF detector response which can be characterised experimentally as an input for simulation. In this study, we assume an ideal detector sensitivity with $D(k)$ equal to 1 for points $k$ on the detector and 0 otherwise in the diffraction space. 

% ! Explain frozen phonon.
Since the incident electrons travel fast as compared to the vibration period of the atoms, the atoms are seen as a frozen snapshot. Therefore, in the frozen phonon approach, the observed electron intensity distribution $|\psi(k, R, z)|^2$ in Eq.~\ref{eq:adf} is calculated for many different atom configurations following the Einstein model and the resulting intensity distributions are averaged over time. The frozen phonon calculations allow us to separate the elastic and thermally scattered electrons. Following Ref.~\cite{VanDyck2009}, the exit wavefunction in reciprocal/real space can be expressed as:
\begin{equation}
    \psi(k/r,\tau) = \langle \psi(k/r,\tau)\rangle +\delta\psi(k/r,\tau),
    \label{eq:exit_wave}
\end{equation}
where $k/r$ is either the reciprocal/real space vector as defined previously, $\tau$ represents a frozen phonon configuration of atom positions, $\langle \rangle$ is the average operation over different phonon configurations and $\delta\psi(k/r,\tau)$ is the deviation from the average wavefunction for a particular phonon configuration. The total intensity $\langle |\psi(k/r,\tau)|^2 \rangle$ is the incoherent sum of electrons averaged over the phonon configurations:
\begin{equation}
    \underbrace{\langle |\psi(k/r,\tau)|^2 \rangle}_{\text{Total}} =\underbrace{|\langle \psi(k/r,\tau)\rangle|^2}_{\text{Elastic}} + \underbrace{\langle |\delta\psi(k/r,\tau)|^2\rangle}_{\text{TDS}}.
    \label{eq:phonon_contribution}
\end{equation}
In this equation, the elastic scattering contribution $|\langle \psi(k/r,\tau)\rangle|^2$ is the modulus square of the averaged wavefunction and the thermal diffuse scattering (TDS) contribution $\langle |\delta\psi(k/r,t)|^2\rangle$ is the average of the modulus square of the wavefunction deviations. When substituting Eq.~\ref{eq:phonon_contribution} in Eq.~\ref{eq:adf}, the elastic and TDS contributions to ADF signal can be separated. For more quantum mechanical approach of treating phonons, the electron intensity can be considered as the incoherent sum of electrons scattered from different initial states of phonons according to their probability distribution, known as quantum excitation of phonon (QEP) ~\cite{forbes2010quantum}. QEP approach is numerically equivalent as the well-known frozen phonon but with different unpinning concept ~\cite{VanDyck2009, forbes2010quantum}. Specifically, for a single electron scattering, QEP considers all phonon configurations. In contrast, frozen phonon treats single electron scattered from only one phonon configuration. QEP can also separate the elastic and thermal diffuse scattering using similar treatment as frozen phonon, see  ~\cite{forbes2010quantum} for details.

The ADF intensities can also be calculated with the absorptive potential approach \cite{Allen1990,Bird1990}. However, an inherent drawback of this approach is that once electrons are absorbed, further elastic or inelastic scattering of the thermally scattered electrons is not accounted for in the simulation and consequently could not properly describe the multiple scattering in a thick sample \cite{Allen2015}. For a detailed comparison study between the incoherent absorptive potential and frozen phonon, see \cite{Alania2018}. Therefore, we will take the frozen phonon (and numerically equivalent QEP) approach in this study.\par

% The wavefunction deviation is averaged to zero as expected.
% \begin{equation}
%     \langle\delta(k/r,t)\rangle = 0.
%     \label{eq:average}
% \end{equation}

% ! Explain about transverse incoherence which enables the use of cross-sections.

Since ADF intensities are dominated by thermally scattered electrons particularly at high angles, which are associated with random phase shifts of transmission functions, one may well suspect that the ADF signal is transverse incoherent due to phonon scattering \cite{Pennycook1990}. Transverse incoherence can be assumed if the image intensity can be written as a convolution of the probe intensity and the object function, which is peaked at the atomic column positions. Transverse incoherence not just yields a directly interpretable image but also allows us to associate the scattered intensities with atomic columns, enabling the quantification of scattering cross-sections. Later analysis \cite{Loane1992,Jesson1993,Nellist1999} showed that phonon scattering is not a prerequisite for transverse incoherent imaging. In fact, transverse incoherence is established due to the geometry of the ADF detector.  The integration over the detector removes the sensitivity to coherent interference effects \cite{nellist2011principles}. However, the detector itself is not efficient in destroying the coherence along the electron beam direction -- which we refer to as longitudinal incoherence -- where phonon scattering will have a more significant effect. Longitudinal incoherence means that the image intensity can be written as a summation of signals generated at each slice along the electron beam. The origin and extent of longitudinal coherence in ADF imaging have not been previously considered.\par

% !Explain about how the EDX yield can be calculated once the fast electron density in the crystal is known.
A fast electron can also excite atomic inner-shell electrons to higher unoccupied states followed by de-excitations via Auger electrons or characteristic X-ray emissions. The EDX effective potential calculates the transition probabilities with all possible energy-momentum transfers and all final continuum states explicitly summed up ~\cite{Oxley2000,Dwyer2013,Allen2015}:
\begin{equation}
    V_{EDX}(r,z) = \frac{\pi m}{h^2} \sum_{n} \frac{1}{k_{n}} |H_{n0}(r,z)\vert^2,
\end{equation}
where $H_{n0}$ is the projected transition matrix element of a electron excited from the initial state $\ket{0}$ to final state $\ket{n}$ with certain energy loss, $k_{n} = \frac{1}{\lambda_{n}}$ is the wave number of the inelastically scattered electron associated with the $\ket{0}$ to $\ket{n}$ excitation. The EDX signal can be considered as the cumulative sum of the probe convoluted with the effective potential at each thickness, resulting an incoherent form for image formation:
\begin{equation}
    I_{EDX}(R) = \frac{4 \pi}{h v}\sum_{z} \int  V_{EDX}(r,z)|\psi(r,R,z)\vert^2 d^2r.
\label{eq:edx}
\end{equation}
$V_{EDX}(r,z)$ is the EDX effective ionisation potential projected for a single plane of atoms at a depth $z$ for a particular X-ray emission. Note that EDX is influenced by dynamical scattering before ionisation with the altered probe intensity convolves with the EDX effective potential. The elastic scattering after ionisation has no further consequences in EDX, which is different from the double channeling situation for EELS. Therefore, the EDX intensities can be written as a summation over the sample thickness for each element at each slice and is longitudinal incoherent. Here, we assume that all excited states for the targeted orbital at the ground state leads to the generation of an X-ray and that the detector reaches the full solid angle. In practice, for full quantification of EDX signals, we should also consider (a) the fluorescence yield of X-rays, (b) the detector geometry, efficiency and shadowing \cite{Xu2016a} and (c) the absorption and scattering of X-rays in their pathway toward the detector \cite{Xu2018}. To simplify the quantification, the effects (a) and (b) simply scales Eq.~\ref{eq:edx} and can be taken into account using the microscope dependent partial cross-section \cite{Varambhia2018}. Absorption (effect (c)) is usually negligible for nanostructured materials due to its small size but should be considered when its effect cannot be ignored in some systems (Ni-Al for example) due to the strong absorption among different elements. One can check the database in \cite{henke1993x} if strong X-ray interaction exists in the system of interest. 

If ADF is also longitudinal incoherent similar to EDX, we can expect a linear dependence of ADF-EDX in the presence of channelling and hence their scattering cross-sections, obtained after integration over the scanned area, will have a linear dependence as well. To investigate the longitudinal incoherence for ADF signals under different conditions, we will compare the channelling behaviour of ADF and EDX in the next section.

In this study, we used muSTEM \cite{Allen2015} to simulate the CBED, ADF and EDX signals for pure elements in Section~\ref{sec:dependence}. To ameliorate the memory requirement, muSTEM augments phonon configurations by random translation of pre-calculated transmission functions by an integer number of unit cells in each direction, which makes it not suitable for non-periodic structures. Since the on-the-fly calculation is not accessible in the current version of muSTEM, a large amount of pre-calculated transmission functions without random phase translation is still doable for small nanoparticles as performed before \cite{MacArthur2017} but not feasible for thick high entropy alloys in this study. Therefore, we take the EDX effective potential based on the inelastic scattering factor tabulated in muSTEM \cite{Allen2015,Oxley2000} and then implemented it in MULTEM \cite{Lobato2015,Lobato2016b} for benchmark in Section~\ref{sec:ALM1.2} and for the high entropy bulk alloys in Section~\ref{sec:ALM1.3}. Note that our EDX implementation is still at the proof-of-concept stage that is not optimised for GPU acceleration. Thus, for small core-shell nanoparticle case studies, we still used muSTEM. We are currently developing our own EDX ionisation potentials from first-principles and GPU implementation of EDX multislice for future studies. 

% Describe that EDX is fully incoherent, so we use a comparison to test longitudinal incoherence in ADF. 
\section{Relationship between ADF-EDX scattering cross-sections}
\label{sec:dependence} 

\begin{figure}[htbp]
    \centering
    \includegraphics[width=1\textwidth]{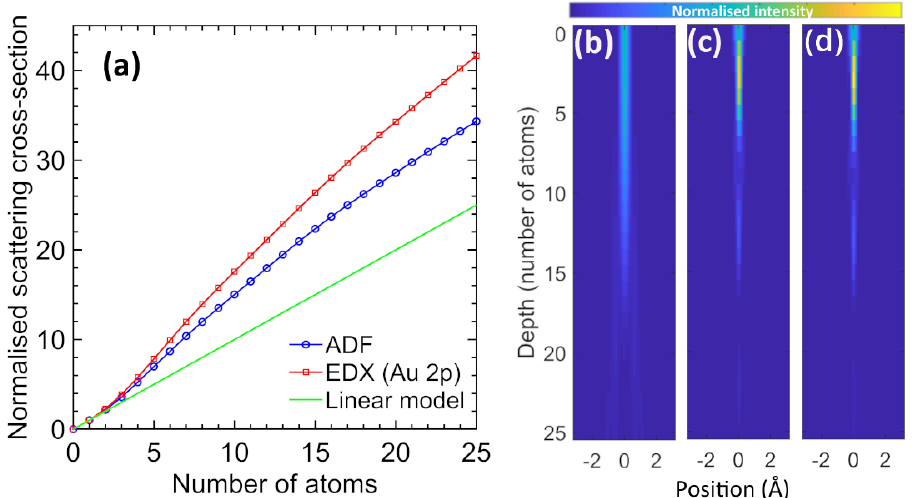}
    \caption{(a) Plots of ADF (with collection semi-angle of 50-150 mrad) and EDX (using transition potential of the 2p orbital, corresponding to the Au L peak) scattering cross-sections as a function of the number of atoms for an Au face-centred cubic crystal in a [1 0 0] direction. The scattering cross-sections are normalised against those of single atoms and compared with the linear model. Cross-sectional depth profile of the electron probability for an aberration-corrected probe in (b) vacuum, (c) a single isolated Au atomic column, and (d) a Au atomic column in a crystal.}
    \label{fig:ADF_EDX_SCS}
\end{figure}
% From the scattering physics, we the linear dependence between ADF and EDX is originated from the incoherent nature of the phonon scattered electrons and X-ray generations. 
ADF and EDX have a non-linear relationship against thickness \cite{VanDenBos2016a} due to dynamical electron scattering, particularly at the atomic scale in zone-axis orientation. This is clear from  Fig.~\ref{fig:ADF_EDX_SCS}(a), where the  ADF and EDX scattering cross-sections are calculated using multislice for a pure Au crystal and normalised against the corresponding values of a single atom. Here we employed a 300 keV aberration-corrected probe with a convergence semi-angle of 20 mrad and ADF collection semi-angle of 50-150 mrad. The detailed settings can be found in Table~\ref{tab1:SimPar} and will be used for following simulations in this study if not stated otherwise. As shown in Fig.~\ref{fig:ADF_EDX_SCS}(a), ADF and EDX scattering cross-sections have a clear deviation from the linear model even for a very thin sample. This can be understood by examining the depth profile of the electron probe free propagation in vacuum and comparing it to that along a single isolated atomic column and an atomic column in a crystal, Fig.~\ref*{fig:ADF_EDX_SCS}(b-d). The presence of atoms focuses the electron probe -- for instance, the probe is narrower with a higher electron density especially for the first few atoms in (c-d) compared to in vacuum (a) in Fig.~\ref*{fig:ADF_EDX_SCS} -- since their positive nuclei act as atomic lenses for the negatively charged electrons, known as electron channelling. A strongly focused probe leads to higher yields of EDX and ADF scattering cross-sections, which varies along the electron beam direction due to dynamic scattering. For a well-separated lattice or more importantly a thin sample, the coupling between columns is not significant so that the electron channelling is largely confined to a single column under probing \cite{Martinez2018a}. This behaviour is therefore similar for the isolated column and the full lattice, as shown in Fig.~\ref*{fig:ADF_EDX_SCS}(c-d). The picture for closely-spaced atomic columns in a thick sample is different since the electron beam may channel, for instance, between the dumbbell structure in Si at larger depths \cite{Hovden2012}. 

\begin{table}
    \centering
    \caption{Settings used for  multislice simulations of different crystals.
    }
    \begin{tabular}{lll}
    \hline
    Multislice settings & Acceleration voltage & $300~\mathrm{kV}$\\
    & Defocus &  $0~\mathrm{nm}$\\
    & Spherical aberration & $0~\mathrm{mm}$\\
    & Convergence semi-angle & $20.0~\mathrm{mrad}$\\
    & Potential pixel size & $4.38~$pm\\
    & STEM image pixel size & $0.24~$\AA\\
    & ADF detector angle & $50-150~$mrad\\[0.2cm]
    Phonon settings  & Number of phonon configurations & $30$\\
    & Al root-mean-square displacement & $0.1012~$\AA\\
    & Ag root-mean-square displacement & $0.0966~$\AA\\
    & Pt root-mean-square displacement & $0.0686~$\AA\\
    & Au root-mean-square displacement & $0.0884~$\AA\\
    \hline
    \end{tabular}
\label{tab1:SimPar}
\end{table}

\begin{figure}[htbp]
    \centering
    \includegraphics[width=0.8\textwidth]{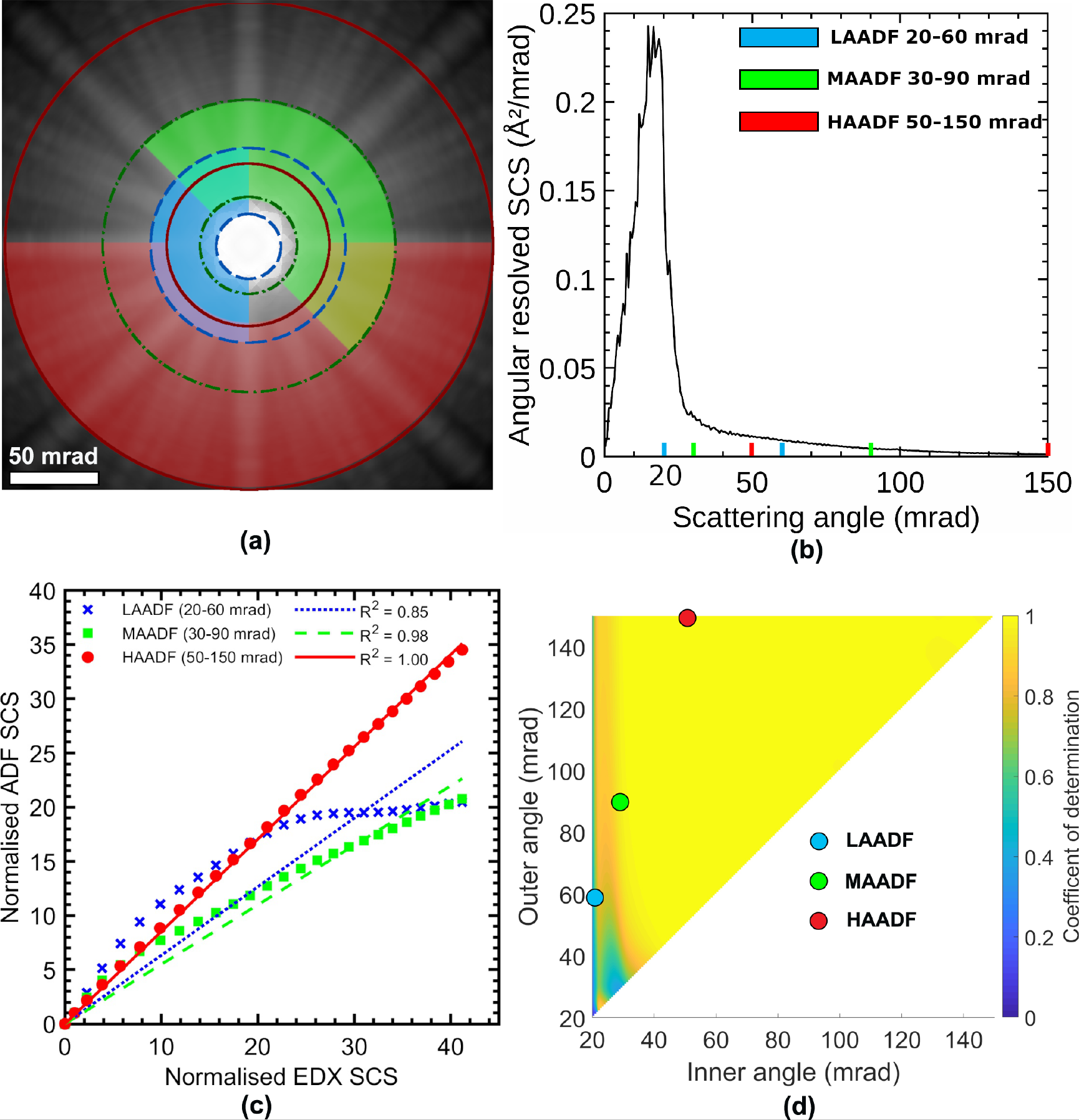}
    \caption{(a) PACBED pattern (shown on a log scale) to demonstrate the range of the LAADF (20-60 mrad), MAADF (30-90 mrad), HAADF (50-150 mrad) detectors. As those detectors overlap,  only half of the detectors are colored for a better visualisation of their collection angles with the other half indicated by solid or dashed lines. (b) Angular resolved scattering cross-section as a function of scattering angle with the ranges for LAADF, MAADF and HAADF highlighted. (c) LAADF, MAADF and HAADF scattering cross-sections as a function of the normalised EDX scattering cross-sections together with a linear regression line. (d) Coefficient of determination $R^2$ of the ADF-EDX linear dependence for a range of different inner and outer collection angles. The simulations were performed for an Au crystal in a [0 0 1] direction with varying thicknesses (1-25 atoms), illuminated using 300 keV electrons with a 20 mrad condenser aperture and no lens aberrations.}
    \label{fig:linearity}
\end{figure}

Although Fig.~\ref{fig:ADF_EDX_SCS}(a) shows that ADF and EDX have a non-linear relationship against sample thickness, we might expect the two signals to follow an identical trend if they are fully incoherent. To test the ADF longitudinal incoherence as a function of scattering angles, we examined the dependence between the two signal modes numerically using multislice calculations. Position averaged convergent beam electron diffraction (PACBED) patterns were computed together with EDX for a unit cell in a pure Au crystal with thicknesses of 1-25 atoms (corresponding to 0-10 nm). By radially integrating a PACBED pattern in the azimuthal direction and dividing by the number of atomic columns in the scanned area, angular resolved scattering cross-sections are obtained, which are then integrated for all possible inner and outer collection angles to obtain the corresponding ADF scattering cross-sections. For instance, three typical ranges for low angle (LAADF 20-60 mrad), medium angle (MAADF 30-90 mrad) and high angle ADF (HAADF 50-150 mrad) are shown in Fig.~\ref{fig:linearity}(a-b). This operation is applied to all PACBED patterns at different thicknesses and the retrieved ADF scattering cross-sections are plotted against EDX scattering cross-sections for the same column thickness in Fig.~\ref{fig:linearity}(c). These ADF and EDX scattering cross-sections are fitted using linear regression. Whereas HAADF has a perfect linear dependence against EDX for different thicknesses, LAADF and MAADF do not show such a relationship. The goodness of fit of the linear regression model can be quantitatively measured by the coefficient of determination $R^2$, which is defined as: 
\begin{equation}
    R^2=1- {\frac{\sum_{i=1}^n (\sigma_{i}-\sigma_{i}^{lin})^2}{\sum_{i=1}^n (\sigma_{i}-\bar{\sigma})^2}},
\end{equation}
with $\sigma_{i}$ the simulated ADF cross-section, $\sigma_{i}^{lin}$ the predicted ADF value based on linear regression, and $\bar{\sigma}$ the mean value of the simulated ADF cross-sections. A perfect linear dependence between the ADF and EDX signals means that the $R^2$ value equals 1. Fig.~\ref{fig:linearity}(d) shows the $R^2$ value as a function of inner and outer detector angle. Since the EDX signal is perfectly incoherent, this graph may be considered as an ADF longitudinal incoherence map. The results reassure our common understanding that the HAADF signal is incoherent while signals recorded at low angles are not. Note that the ADF coherence measured in this approach depends on the sample and microscope parameters. For instance, an ADF detector being incoherent for a thin sample with light elements may become semi-coherent for a thick sample with heavy elements.\par

\begin{figure}[htbp]
    \centering
    \includegraphics[width=1\textwidth]{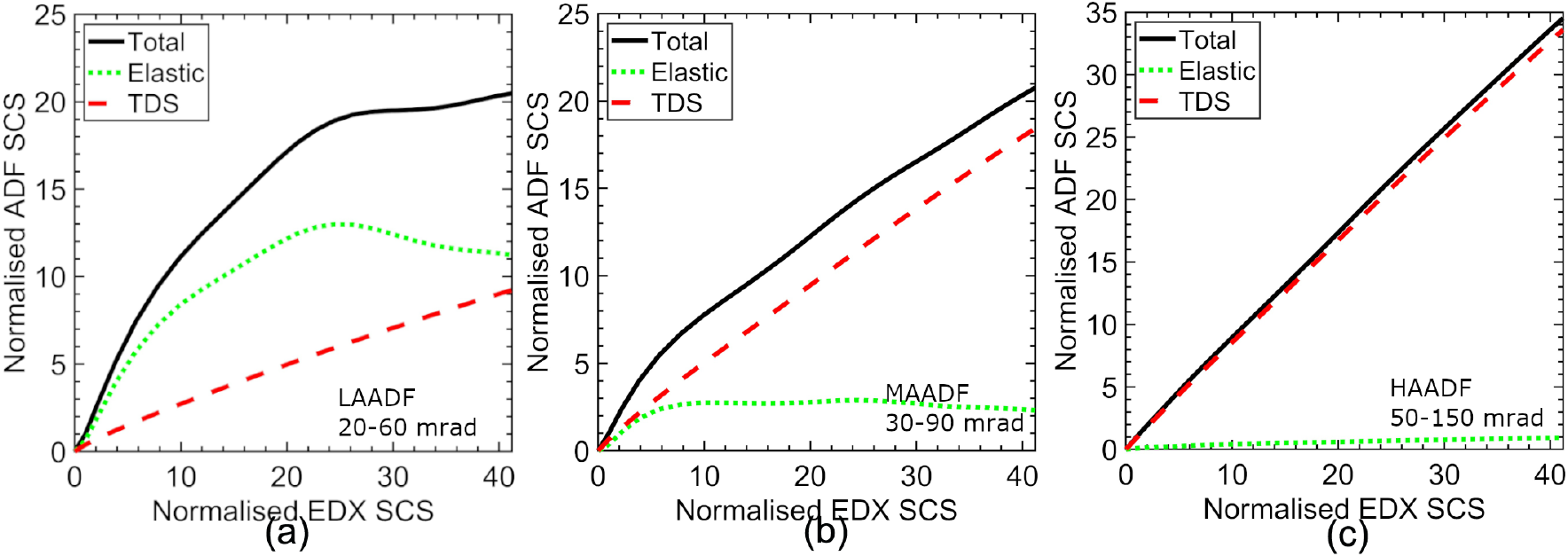}
    \caption{Plots of normalised ADF scattering cross-sections against EDX scattering cross-sections for (a) LAADF, (b) MAADF and (c) HAADF.}
    \label{fig:elastic_contributions}
\end{figure}

To understand the deviation of the ADF signal from perfect incoherence at low and medium angles, we can separate the contributions of elastic scattering and thermal diffuse scattering in the diffraction patterns according to  Eq.~\ref{eq:phonon_contribution}. As shown in Fig.~\ref{fig:elastic_contributions}(a-b), the elastic signal has a significant contribution at low and medium angles of the ADF detector resulting in a deviation of the linearity against EDX. In contrast to the elastic contribution, phonon scattered signals are almost linear against EDX with increasing thickness and dominate the HAADF intensities as shown in, Fig.~\ref{fig:elastic_contributions}(c). 

\begin{figure}[htbp]
    \centering
    \includegraphics[width=1\textwidth]{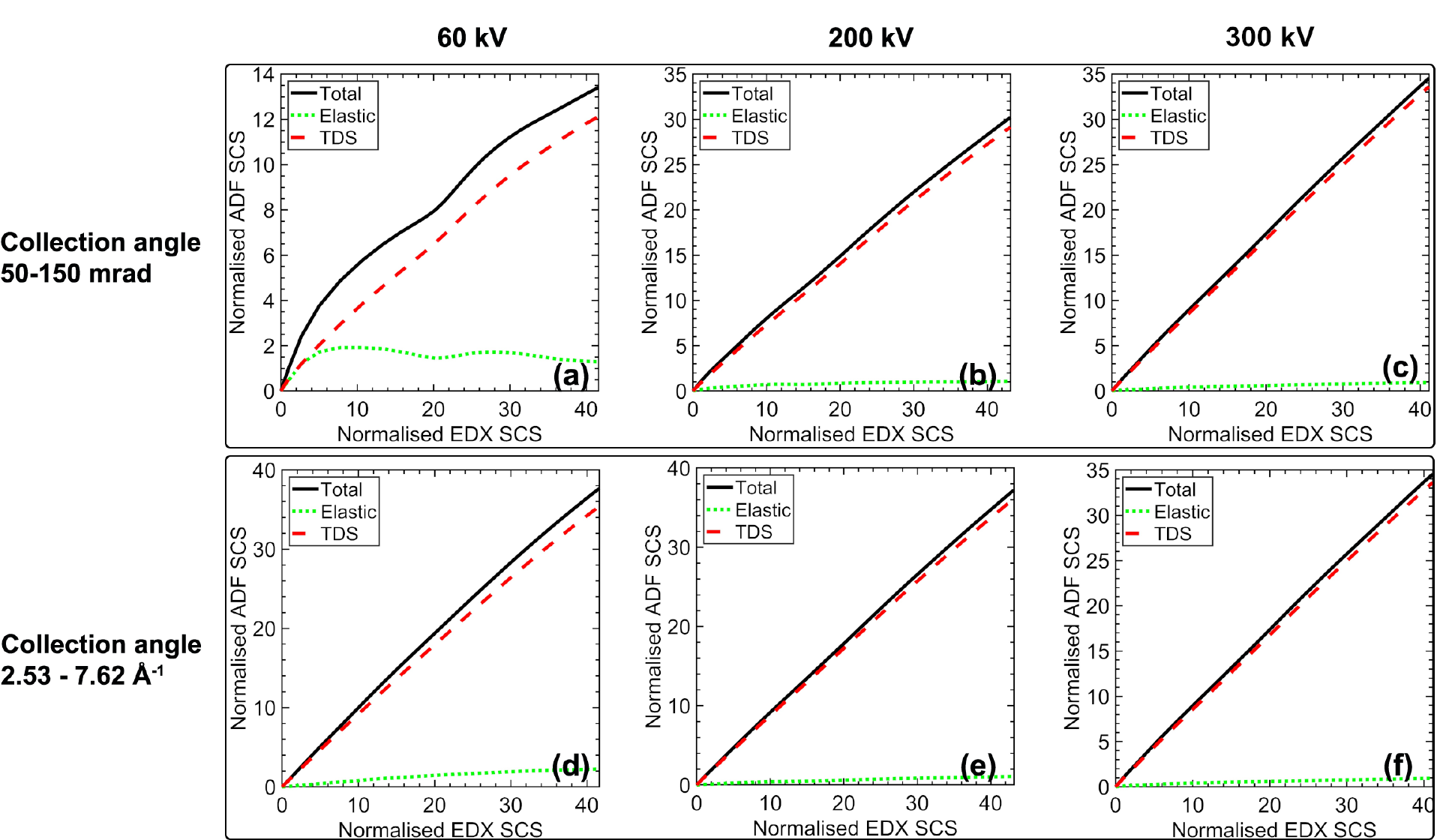}
    \caption{Plots of normalised ADF scattering cross-sections against EDX scattering cross-sections for different acceleration voltages with (a-c) the same collection angle in mrad; (d-f) the same collection angle in 1/\AA.}
    \label{fig:changing_voltage}
\end{figure}

To investigate the longitudinal incoherence with varying voltage, the ADF collection range is measured in terms of the scattering vector in $\angstrom^{-1}$ and the geometric angle in mrad. As shown in Fig.~\ref{fig:changing_voltage}, the ADF-EDX linear dependence of conventionally considered HAADF angle (50-150 mrad) at 300 kV could break down at 60 kV. In contrast, the linearity can be well-kept when we translate the collection angle of 50-150 mrad to 2.53-7.62 $\angstrom^{-1}$ at 300 kV and apply it for a lower voltage. The reason is that the positions of CBED disks for elastic scattering is controlled by the lattice spacing while the phonon scattering is characterised by root-mean-sqaure displacement of the element, both are constant measured by the $\angstrom^{-1}$ in the diffraction plane. Further angular resolved scattering cross-section calculations show that the range of where the thermal diffuse scattering starts to dominate is relatively invariant to the acceleration voltage. Though for the case of 60 kV (2.53-7.62 $\angstrom^{-1}$ or equivalently 123-370 mrad), ADF scattering cross-sections have a small but not negligible contribution from elastic signals, its relationship against EDX is still linear. The elastic contribution in this case is due to first order Laue zone, which falls within ADF range at low voltage. 

In this section, we showed that the integration over the ADF detector, which destroys the transverse coherence, does not control the longitudinal coherence. One must select a sufficiently high inner collection angle to make the truly incoherent phonon scattered electrons dominate the ADF signal. Note that in this simulation study, we followed the conventional uncorrelated Einstein model of phonons generation that displaces atoms in 2D. The probability distribution $P(\tau)$ depending on its position $\tau$
\begin{equation}
    P(\tau) = \frac{1}{\sqrt{2\pi\langle u^2 \rangle}} exp[\frac{(\tau - \tau_0)^2}{\langle u^2 \rangle}],
\end{equation}
where $\langle u^2 \rangle$ is the mean squared displacement of the atom. The root-mean-square displacement $\sqrt{\langle u^2 \rangle}$ for different elements used in this study are given in Table~\ref{tab1:SimPar}. The proper 3D phonon with realistic dispersion, which includes a greater excitation of long-wavelength correlated phonons, is beyond the scope of this study.

\section{Extending the atomic lensing model for spectroscopy}
\label{sec:ALM}
In the previous section, we examined a robust linear dependence between EDX and ADF via multislice simulations. However, we should note that such a simulation is computationally expensive. For a 20-atom-thick binary alloy, there are more than 1 million different 3D column configurations to cover the entire composition range. The situation is even worse when the number of elements further increases. Therefore, to quantify EDX at atomic resolution, a fast prediction method is needed for the elemental quantification taking dynamical diffraction into account. The atomic lensing model, which is a non-linear model under channelling conditions, was previously developed for ADF and successfully applied in atom counting of mixed columns in an Au@Ag core-shell nanoparticle \cite{VandenBos2019a,VanDenBos2016a}. Based on the incoherent imaging of ADF and EDX signals, one would expect that this model also works for EDX. In section~\ref{sec:ALM1.1}, the theoretical extension of the atomic lensing model to EDX is described. Section~\ref{sec:ALM1.2} will benchmark the computational complexity, speed and accuracy of the atomic lensing model compared to the multislice and the recently developed PRISM algorithm \cite{ophus2017}. Then, in Section~\ref{sec:ALM1.3}, we will apply the atomic lensing model to some challenging systems including a core-shell nanoparticle and high entropy alloy and will compare the predictions against the results from multislice simulations to showcase its advantages and limitations. 

\subsection{Channelling theory of atomic lensing model for spectroscopy} 
\label{sec:ALM1.1}
If we assume that the electron probe wavefunction stays constant in the crystal with respect to thickness and that the scattering from each atom can be considered as being incoherent with respect to other atoms, the scattering cross-section is a simple addition of the effective potentials. The scattering cross-section will then increase linearly against sample thickness, noted as the linear incoherent model. In reality, the electron wave function scatters dynamically giving varying contributions at different depths and hence making elemental quantification difficult. In this section, we will expand the atomic lensing model developed previously for ADF \cite{VandenBos2019a,VanDenBos2016a} to spectroscopy with a simple modification. In the atomic lensing model, we treat dynamical scattering as a superposition of individual atoms focusing the incident electrons. Here, we assume that the electron channelling effect of these individual columns alters the electron probe function and that the cross-talk of surrounding columns is negligible. By comparing the electron probe profile as a function of depth down an isolated column and an atomic column in a crystal shown in Fig.~\ref{fig:ADF_EDX_SCS}(c-d), the dynamical scattering is indeed largely confined to the individual columns for a sufficiently thin crystal if columns are well-separated. Following the derivation given in \cite{VandenBos2019a}, the focusing effect of an atomic column is given by
\begin{equation}
    F_{col}(1 \rightarrow n)=\frac{1}{\Theta_{col, Z(n+1)}(1)} \frac{d \Theta_{col}}{d n}=\frac{\Theta_{col}(n+1)-\Theta_{col}(n)}{\Theta_{col, Z(n+1)}(1)},
\end{equation}
where $F_{col}(1 \rightarrow n)$ is the focusing effect of a column of n atoms, with atoms located at the 1st to nth position.  $\Theta_{col}(n)$ is the scattering cross-section of a column consisting of n atoms. The difference between the scattering cross-section of n+1 atoms and n atoms is normalised by that of a single atom ${\Theta_{col, Z(n+1)}(1)}$ to measure the non-linear contribution from the (n+1)th atom due to the lensing effect of the previous n atoms, where $Z(n+1)$ is the type of element for the (n+1)th atom. The lensing effect of an individual atom can be determined from the superposition principle:
\begin{equation}
    L_{Z}(n)=\frac{d F_{col}}{d n}=F_{col}(1 \rightarrow n)-F_{col}(2 \rightarrow n),
\end{equation}
where $L_{Z}(n)$ is the lensing factor of the 1st atom with atomic number Z on the (n+1)th atom. Similar as in optics, the lensing effect $L_{Z}(n)$ only depends on the relative distance away from this atomic lens, not its absolute position \cite{VandenBos2019a}. For instance, the lensing effect of the 1st atom on the nth atom is equal to that of the 2nd atom on the (n+1)th atom (if we simply shift the absolute position while the atoms are the same). Therefore, though the scattering cross-section is non-linear against the sample thickness due to channelling, its second derivative can be linearly additive.

Following the superposition of lensing factors of each individual atom, which can be calculated from pure element libraries, we may predict the scattering cross-section of a mixed column in any ordering. For ADF-STEM, the predicted scattering cross-section is given by ~\cite{VandenBos2019a,VanDenBos2016a}:
\begin{equation}
    \Theta_{col}^{ADF}(N)=\Theta_{col}^{ADF}(N-1)+\left(1+\sum_{n=1}^{N-1} L_{Z(n)}^{ADF}(N-n)\right) \Theta_{col,Z(N)}^{ADF}(1),
\label{eq:alm_adf}
\end{equation}
where $Z(n)$ is the atomic number of the nth atom in a mixed column. The lensing factor $L_{Z}(n)$ of each atom of a column alters the electron probe function, yielding a non-linear response due to channelling, which is summed to predict the focusing effect for the next atom in sequence. The resulting scattering cross-section $\Theta_{col}^{ADF}(N)$ is predicted for a mixed column at the depth of N atoms. 

For spectroscopy being an incoherent imaging technique, the scattering cross-section for each element can be written as:
\begin{equation}
    \Theta_{col}^{Spec}(N,Z(N))=\Theta_{col}^{Spec}(N-1,Z(N))+\left(1+\sum_{n=1}^{N-1} L_{Z(n)}^{Spec}(N-n)\right) \Theta_{col}^{Spec}(1,Z(N)),
\label{eq:alm_spec}
\end{equation}
where $\Theta_{col}^{Spec}(N,Z(N))$ is the scattering cross-section matrix of a mixed column with prediction value at the depth of N atoms and element with atomic number Z(N). Note that the atomic number Z(N) is a function of depth and encodes the ordering and number of atoms in a column. The spectroscopy scattering cross-section matrix $\Theta_{col}^{Spec}(N,Z(N))$ is calculated in a step-wise manner, with rows representing the depth and columns representing different elements. For instance, the scattering cross-sections at the Nth row are derived from the (N-1)th row with the increment of cross-section of the element with atomic number Z(N) following the lensing rule. In practice, this requires simulations of the EDX signals for each element to predict the EDX of mixed columns. This will be applied in the Au@Pt core-shell nanoparticle case in Section~\ref{sec:ALM1.3}.

Since there is a strong linear dependence of ADF-EDX as examined in Section~\ref{sec:dependence}, we can also make EDX predictions from ADF: 
\begin{equation}
    \Theta_{col}^{Spec}(N,Z(N))=\Theta_{col}^{Spec}(N-1,Z(N))+\left(1+\sum_{n=1}^{N-1} L_{Z(n)}^{ADF}(N-n)*K(Z(N))\right) \Theta_{col}^{Spec}(1,Z(N)),
\label{eq:alm_spec_adf}
\end{equation}
where $L_{Z(n)}^{ADF}(N-n)$ is the lensing factor resulting from ADF libraries of pure elements and $K(Z(N))$ is the slope of the ADF-EDX linear dependence for the element of interest $Z(N)$. This means that one can use the ADF signals to evaluate the lensing effect of each element, which can be applied to predict EDX scattering cross-sections. The benefit of this approach is not in saving computation time for preparing a particular library. In fact, simulating ADF and EDX takes the same amount of time and one has to simulate EDX to evaluate the ADF-EDX slope anyway. However, the linear ADF-EDX relationship makes it easy to perform empirical simulations of EDX scattering cross-sections based on experimental ADF-EDX data from which the slope of the linear dependence can be determined. It also enables the transfer of a trained ADF scattering cross-section prediction neutral network (not yet published) to EDX without retraining the network. To test Eq.~\ref{eq:alm_spec_adf}, we calculate the full ADF library at each thickness and EDX library at a finite number of thicknesses to retrieve the ADF-EDX slope using frozen phonon calculations for the high entropy alloy case in Section~\ref{sec:ALM1.3}.

% Finally, we note that the atomic lensing model should also be applicable for EELS with a large collection angle and inner-shell excitation. The EELS imaging is non-local and could be strongly influenced by the elastic scattering after ionisation before reaching the detector. Double channelling calculation  -- considering the multiple elastic scattering before and after the inelastic excitation event -- is needed to properly describe the observed inelastic electron density \cite{Verbeeck2009,Dwyer2008}. However, multislice simulations have shown that the local approximation (with longitudinal incoherent imaging equation similar to Eq.~\ref{eq:edx}) could hold for a sufficiently high inner collection angle \cite{Dwyer2005}. In addition, a combination of experiments and simulations have verified that one can correct the elastic artefact in EELS intensities by the simultaneously-acquired incoherent bright filed \cite{Zhu2013a}, where the corrected map shows similar contrast as incoherent calculations. A qualitative comparison between experiments and simulations suggested a good agreement for K- and L-shells while M-shells are significantly overestimated \cite{Zhu2014a}. To apply the atomic lensing model for EELS if the longitudinal incoherent imaging could be ensured, one can simply use the EELS scattering cross-section in Eq.~\ref{eq:alm_spec}. The comparison of the atomic lensing model predicted EELS against double channelling simulations is beyond the scope of this paper.

\subsection{Computational complexity and accuracy}
\label{sec:ALM1.2}

% computational complexity of multislice, PRISM and ALM
A major challenge for spectroscopy quantification of complex nanostructures is to consider the channelling effect in mixed columns. The number of possible combinations in the ordering of atoms exceeds the capability of multislice calculations. Recent developments with the PRISM algorithm provides a significantly speedup alternative \cite{ophus2017,pryor2017}, which is now available for both STEM \cite{pryor2017,dacosta2021,madsen2021} and EELS \cite{brown2019} simulations. PRISM combines the Bloch wave and multisclice via the scattering matrix to alleviate the repetitive computation cost involved in each scanning probe positions \cite{ophus2017}. This is particularly attractive in case of a large field of view. The accelerated speed is at the cost of accuracy \cite{ophus2017,dacosta2021,pelz2021}. However, when facing the ordering possibilities for each column multiplied by the number of columns that are potentially mixed, the PRISM algorithm can also be time consuming. In contrast, the atomic lensing model is a column-by-column prediction framework \cite{VandenBos2019a,VanDenBos2016a}, which might be less accurate but provides a much faster albeit rough estimation. In this section, we will examine the computational cost and accuracy of atomic lensing model against multislice calculations so that one can make a rational choice. We also include  PRISM algorithm in the computational cost benchmark as an alternative option.   

Here we follow the analysis in \cite{ophus2017} to make an estimate of the calculation time. The computational complexity for each algorithm is given in Table~\ref{table:complexity} together with the parameters used. In contrast to the previous analysis, we also take into account the number of phonon configurations and the number of column ordering configurations, as they are indeed just common multiplication factors for multislice and PRISM but not for the atomic lensing model. For the multislice algorithm with a supercell sampled by $N \times N$ pixels, each slice requires 5 forward and backward Fourier transformations (complexity: $5Nlog_2{N}$) together with a wave function multiplication with the potential in real space and with the Fresnel propagator in reciprocal space (complexity: $2N^2$) \cite{ophus2017}. This complexity is amplified with (1) the number of slices $H$, (2) the number of probe positions $P$, (3) the number of phonon configurations $T$ and (4) the number of possible orderings $O$ in mixed columns. The PRISM algorithm only needs to perform the repetitive transmission-propagation in the multisclice once to construct the scattering matrix for each parallel beam sampled. The number of beams needed $B$ can be factorised by the interpolation factor $f$. The effect of the number of probe positions $P$ is added later, which is outside of the multislice loop (complexity: $PBN^2/4f^4$) \cite{ophus2017}. However, the computational time still scales with the ordering possibilities. In contrast, the atomic lensing model only needs the multislice calculations to build the pure element libraries. The following calculations to generate the scattering cross-sections for a mixed column for any ordering are simple numerical operations in Eq.~\ref{eq:alm_adf}-\ref{eq:alm_spec} and are only dependent on the number of possible elements $E$ and the number of atoms (at same order as number of slices $H$) in a column. Note that the scattering cross-section is a single value predicted for a column instead of a full image simulated in multislice and PRISM. Also note that the atomic lensing model prediction for each column is treated totally independent. Hence the total number of orderings for a system is a summation of the orderings in each column. The column-by-column approach simplifies the exploration of ordering and provides a significant speedup in predictions, which however is also the major source of error as we can see later in the benchmark and case studies.

% Also note that the number of ordering configurations for mixed columns is the product of the number of orderings for each columns for multislice and PRISM as in a supercell. 
\begin{table}[!ht]
    \centering
    \caption{Computational complexity of the multislice simulation, the PRISM simulation and the atomic lensing model.}
    % \resizebox{\textwidth}{!}{
    \begin{tabular}{l l}
    \hline
        Algorithm & Computational complexity\\ \hline
        Multislice & $OTHP[5Nlog_2{N}+2N^2]$\\ 
        PRISM & $OT[\frac{HB}{f^2}[5Nlog_2{N}+2N^2]+\frac{PBN^2}{4f^4}]$ \\
        Atomic lensing model & $ETHP[5Nlog_2{N}+2N^2]+OHE$\\\hline
        Parameter & definition \\ \hline
        O &  number of ordering configurations \\
        T &  number of phonon configurations \\
        H &  number of slices \\
        P &  number of probe positions \\
        N &  side length (in pixels) for supercell sampling \\
        B &  number of beams \\
        f &  interpolation factor \\
        E &  number of elements in the system\\
        \hline
    \end{tabular}
    % % \begin{flushleft}
    % \end{flushleft}
    \label{table:complexity}
\end{table}

\begin{figure}[htbp]
    \centering
    \includegraphics[width= 0.4\textwidth]{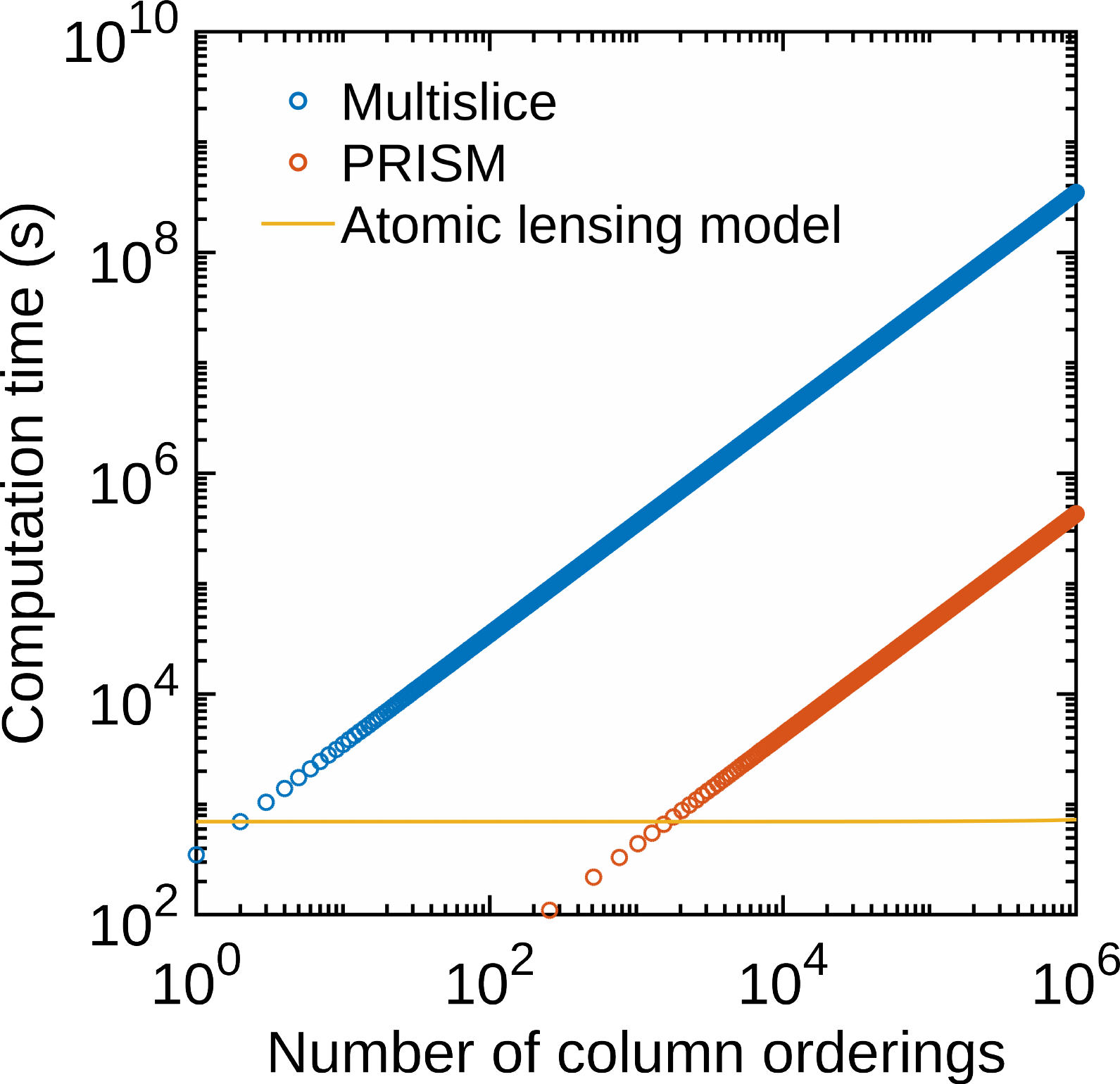}
    \caption{Comparing the computation time for the multisclice simulation, the PRISM simulation and the atomic lensing model for predicting the scattering cross-section against the number of ordering configurations in an Al-Ag binary alloy crystal. }
    \label{fig:time_benchmark}
\end{figure}

\begin{figure}[htbp]
    \centering
    \includegraphics[width= 1\textwidth]{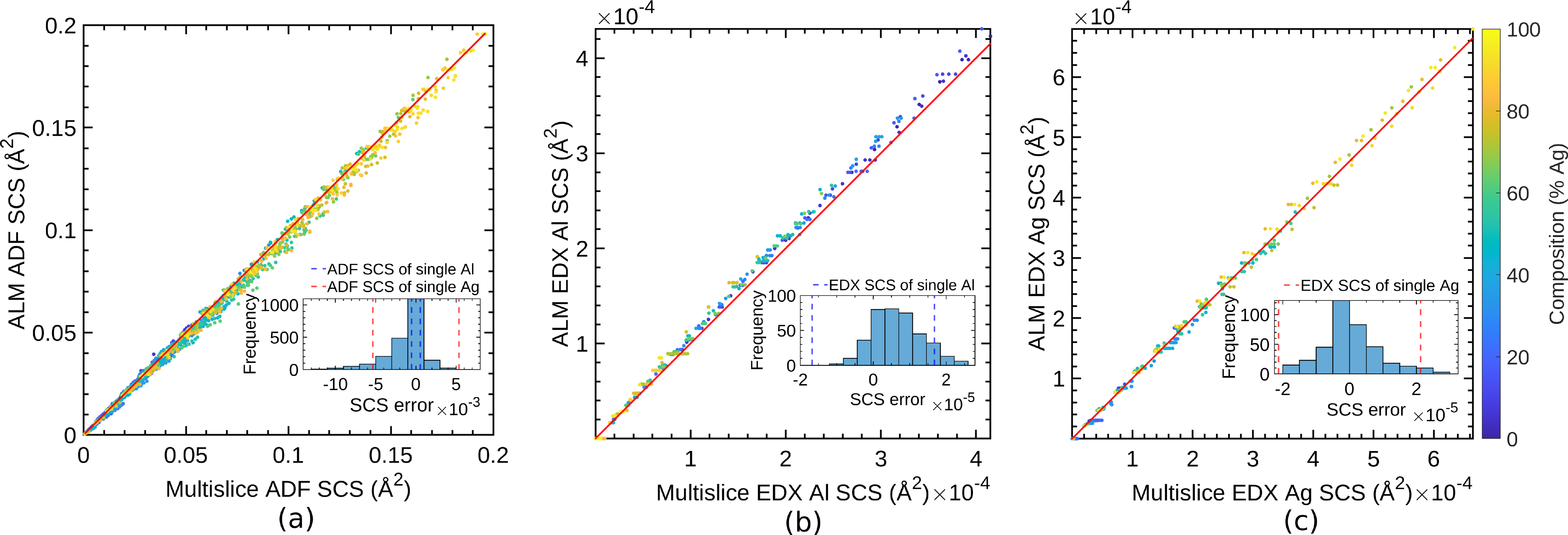}
    \caption{Multisclice simulated against the atomic lensing model predicted scattering cross-sections for (a) ADF, (b) EDX Al and (c) EDX Ag, with a red line indicating the perfect predictions. The histograms of the absolute errors are given in the insets.}
    \label{fig:accuracy_benchmark}
\end{figure}

% Speed and accuracy benchmark
To benchmark the speed and accuracy, we tested the computation time against the number of column orderings in a Al-Ag binary alloy crystal with a random ordering and a supercell made of $8\times8\times20$ face-centred cubic unitcells. We used the MULTEM software \cite{Lobato2015} for the multisclice simulation with the parameters in Table~\ref{tab1:SimPar} and the abTEM software \cite{madsen2021} for the PRISM algorithm with an interpolation factor of 20 tested on a desktop with an Intel i7-8700K CPU and a Nvidia RTX 1080 GPU. We only benchmarked the ADF computation time, because PRISM does not have the EDX capability yet and our prototype EDX multislice is not optimised for GPU (to be implemented). The EDX computational time will be on a similar scale as ADF once optimised. As shown in Fig.~\ref{fig:time_benchmark}, a new multislice simulation is needed for each different ordering, hence its computational time is extrapolated linearly against the number of column orderings to be computed, with each column taking $\sim$ 350 s. The PRISM algorithm outputs all the columns in the input supercell simultaneously thanks to the shared scattering matrix, which is much faster per column ($\sim$ 110 s for 256 columns), but still has a linear scaling against the number of column orderings. In contrast, the most time consuming part of the atomic lensing model is the library generation via multisclice simulations which scales with the number of elements in the system. The prediction, however, is as fast as 29$\pm$5 $\mu$s per column showing an almost constant behaviour in the log-log plot in Fig.~\ref{fig:time_benchmark}. In fact, the atomic lensing model is the only feasible approach that can explore all the ordering possibilities for a 20-atom-thick binary alloy column, taking $\sim$ 30 s to loop over 1 million orderings. Instead of making new predictions again for another column, one can simply adopt the existing predictions as a look-up table for different thicknesses and orderings. Storage of such database increases linearly with the ordering configurations which will eventually become challenging for thick samples.

In order to benchmark the the accuracy, we sampled the Al-Ag alloy composition in the range of 1-99\% Ag with 1\% interval for ADF and 5-95\% Ag with 5\% interval for EDX with different ordering in all columns for each composition. In each case, one column was selected for the probe to scan over the corresponding Voronoi cell and measure its scattering cross-section.  Fig.~\ref{fig:accuracy_benchmark} shows the atomic lensing model predicted ADF and EDX scattering cross-sections against those quantified from multisclice for different thicknesses and compositions (indicated by colors). We can see that most of the predicted values are in close agreement with simulations where the red line indicates a perfect match. The histograms of the absolute errors, defined as the difference between the predicted and simulated values are shown in the insets of Fig.~\ref{fig:accuracy_benchmark}. From these histograms it follows that most of the prediction errors are less than the scattering cross-section of single atoms. We do not compare the PRISM accuracy further in this paper as it has been discussed in several studies \cite{ophus2017,dacosta2021,pelz2021}, which is highly dependent on the interpolation scheme. The interpolation factor of 20 used here corresponds to $\sim$10\% error in PRISM as shown in \cite{ophus2017}. 

% Further analysis (see Fig.S1(a-b)) shows that the ADF mean relative error averaged over all compositions is ~3\% for different thicknesses. The worst case relative error of over 30\% can occur for thin samples because the scattering cross-section itself is small for few atoms, while the absolute errors increase with increasing thickness with the cross-talk contributions from neighbouring columns. Similar trends are also observed for EDX absolute errors with increasing thickness in Fig.S1(c-d). Interestingly, the atomic lensing model seems to slightly overestimate the Al EDX cross-sections with the strong focusing effect of Ag contributing to the increased EDX signals of following Al atoms in the sequence, while the libraries assume the same neighbouring atoms in the presence of beam spreading. The relative errors of EDX are not computed since a column can still have a weak EDX signal of an element even if it is not present in the column due to beam spreading, in which case the atomic lensing model would predict a zero value of the scattering cross-section of this element in EDX and the relative error is 100\%. 

\subsection{Case studies: core-shell nanoparticle and high entropy alloy} 
\label{sec:ALM1.3}
The atomic lensing model allows for a fast generation of scattering cross-sections with the ordering of elements taken into account under the channelling condition. In this section, we will demonstrate the accuracy and limitation of the atomic lensing model in predicting the ADF-EDX scattering cross-sections of mixed columns. The results will be compared against multislice simulations and the linear model. Note that the linear incoherent model here refers to cross-sections increasing linearly with the number of atoms, which is different from the linear dependence between ADF-EDX signals.

\begin{figure}[htbp]
    \centering
    \includegraphics[width= 1\textwidth]{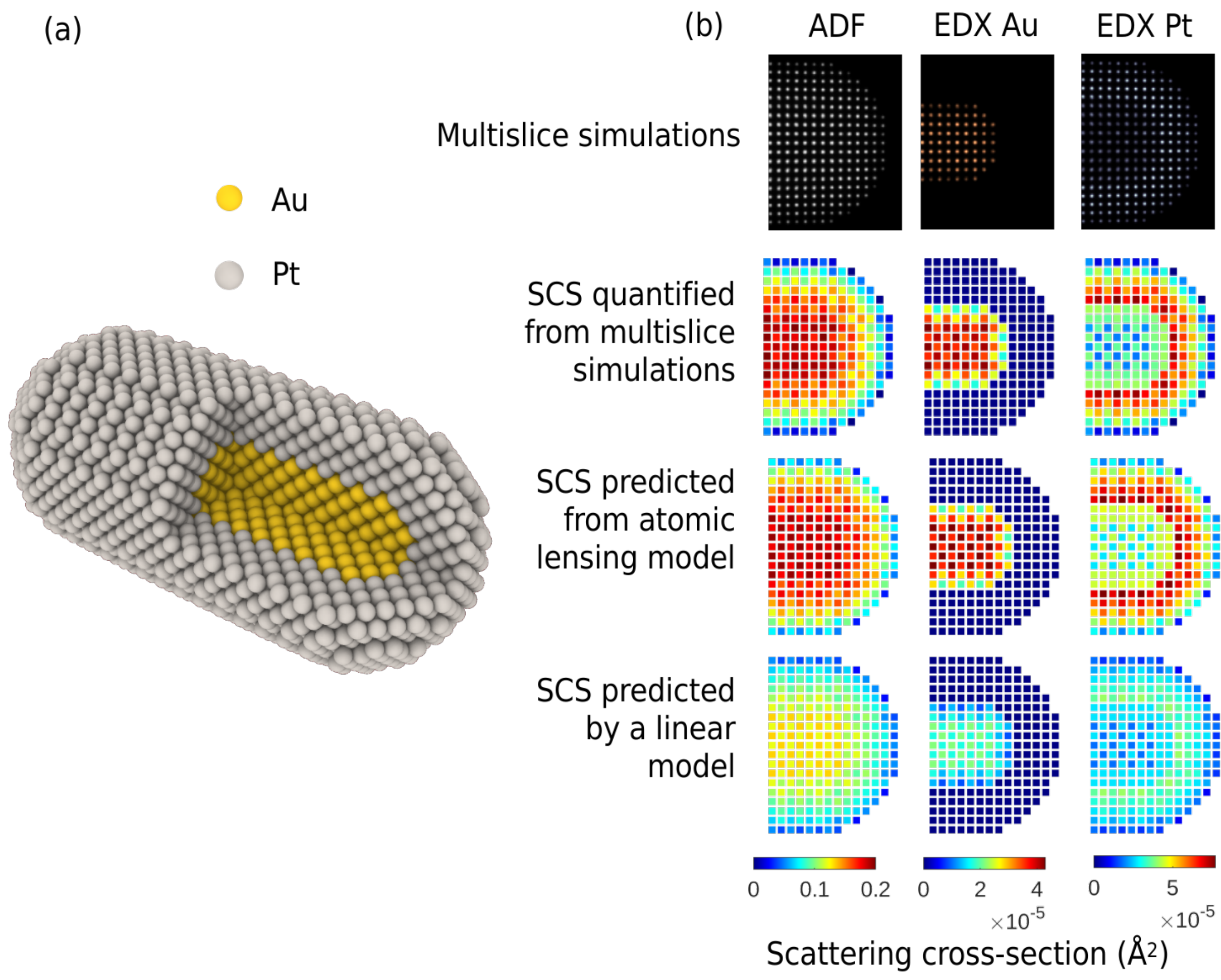}
    \caption{(a) Atomic model of the Au@Pt core-shell nanoparticle. (b) Comparison of the simulated multislice quantified, atomic lensing model (ALM) and linear incoherent model predicted ADF-EDX scattering cross-sections (SCS). The simulation parameters are given in Table~\ref{tab1:SimPar}.}
    \label{fig:core-shell}
\end{figure}

One cannot readily distinguish the presence between Pt and Au based on an ADF image since their atomic numbers only differs by 1. However, we can separate them unambiguously based on their spectroscopy signals as shown in Fig.~\ref{fig:core-shell} for a core-shell Au-Pt nanorod. To quantify the images, both the ADF and EDX scattering cross-sections are extracted from the simulations using Voronoi cell integration, which agree reasonably well with the atomic lensing model predictions. Further relative error analysis shows that the atomic lensing model predictions match nicely (error $<$ 5\%) with most sites for the ADF signal except for those columns at the edges near the vacuum. The EDX signals, however, are systematically overestimated for Au at the core-shell interface and underestimated for Pt in the core-shell region (error $\sim$ 10\%). Those results can be understood from the fact that the atomic lensing model is based on pure elemental libraries, which unavoidably treats the contributions of surrounding columns as pure elements thus deviating from reality. In contrast, the linear model significantly underpredicts the signals since electron channelling is ignored. We noticed that the nanoparticle can undergo surface relaxation leading to misalignment of atomic columns and hence cause a larger error for the atomic lensing model which is based on perfect crystal libraries. In addition, microscopy experiments are often under limited doses thus affecting the measurement accuracy while simulations shown here are at infinite dose. The discussion of atomic lensing model for combined ADF-EDX atom counting with limited dose and simulated particle relaxation is included in a separate paper. 

\begin{figure}[htbp]
    \centering
    \includegraphics[width= \textwidth]{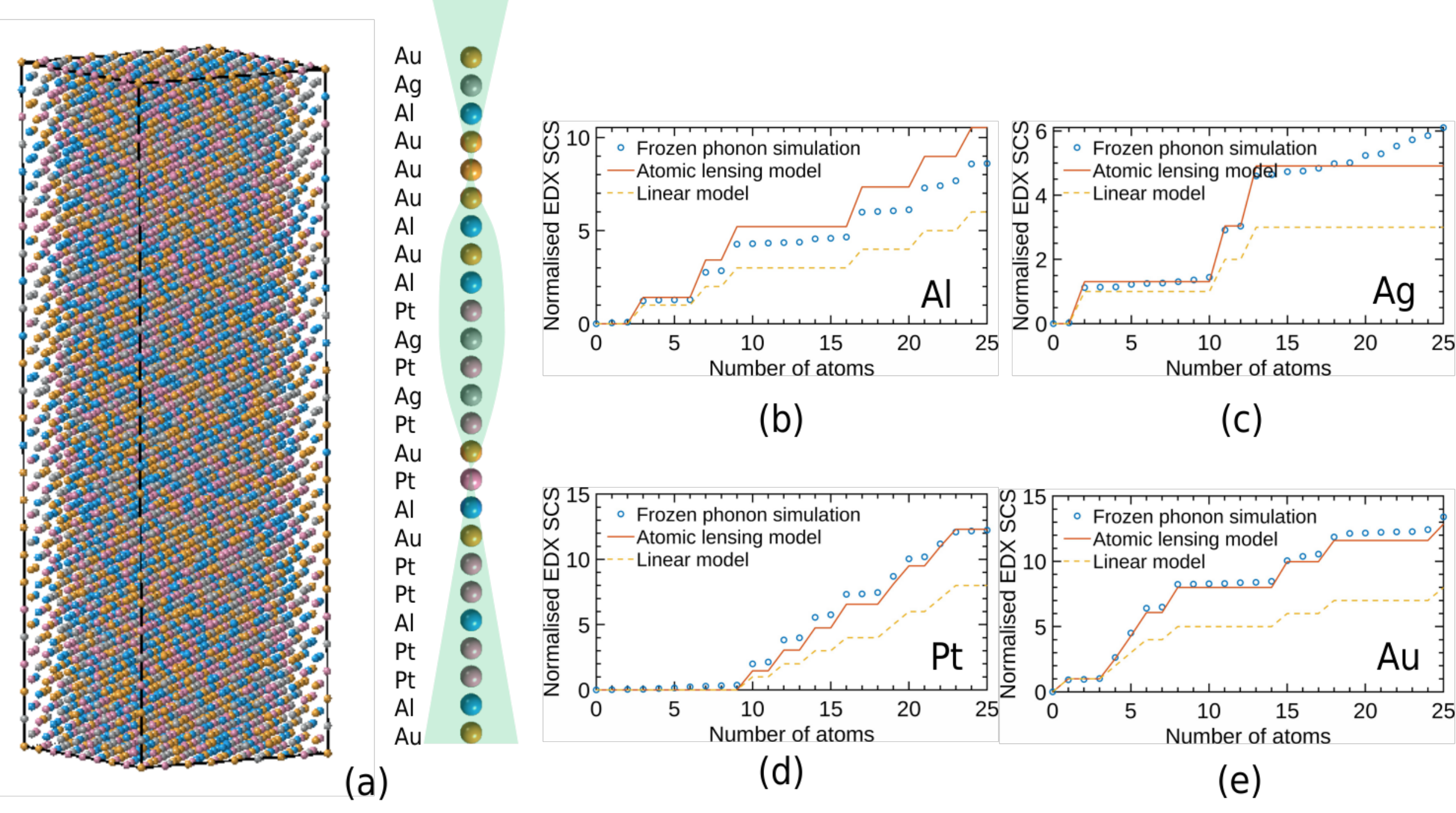}
    \caption{(a) A 3D model of the Al-Ag-Pt-Au high entropy alloy slab with 25 atoms in each atomic column along the electron beam direction. The ordering of a  particular column is given, which is used for comparing the simulated values and predictions from the atomic lensing model and linear model. The normalised EDX scattering cross-sections of this column are plotted as a function of the number of atoms for (b) Al, (c) Ag, (d) Pt and (e) Au respectively. The simulation parameters are given in Table~\ref{tab1:SimPar}.}
    \label{fig:high_entropy}
\end{figure}

To evaluate the atomic lensing model in nano-materials that contain both heavy and light elements which result in complicated electron channelling, we randomly substitute an Au crystal with Al, Ag and Pt, each taking 25\% of the sites of the full lattice, to form a high entropy alloy. The full 3D crystal model and the ordering for a particular column under investigation is given in Fig.~\ref{fig:high_entropy}(a). In Fig.~\ref{fig:high_entropy}(b-e), it is shown that the predicted EDX scattering cross-sections for those elements are in good agreement with simulated results indicating that electron channelling is well captured by the atomic lensing model. However, Fig.~\ref{fig:high_entropy}(c) shows an increasing Ag scattering cross-section against sample thickness, while there is no Ag in the ordering of this column beyond a depth of 13 atoms. The deviations are caused by ignoring the contribution from the Ag atoms in the neighbouring atomic columns. Since the spatial spread of the electron beam increases with increasing thickness, both due to the geometry of a cone-shaped beam and the scattering by the atoms, the EDX contribution from neighbouring atoms will become important and column-by-column analysis may eventually break down. We should be aware of this effect since it is already observed at a thickness of ~15 atoms as shown in Fig.~\ref{fig:high_entropy}(c). A full multislice simulation is then required for each specific case to consider the effect of beam spreading which is beyond this paper. We refer interested readers to \cite{MacArthur2021}  for an example of the quantification of an  heterophase interface. For future studies, we will explore the possibilities of a "hybrid" strategy for the quantification of mixed columns: i.e. using the atomic lensing model to provide good starting predictions, which can then be further refined using multislice or PRISM calculations. 

\section{Conclusions}
In this manuscript, we proposed a method for a fast prediction of the ADF-EDX scattering cross-sections under channelling conditions. EDX signals are fully incoherent following the inelastic scattering theory. For ADF with a sufficiently high inner collection angle, the incoherent phonon scattered electrons dominate the contrast while the elastically scattered electrons also become longitudinal incoherent, thus establishing a linear dependence between ADF and EDX signals against sample thickness. We examined the validity of this linear dependence as a function of ADF collection angles under different microscope conditions. In addition, this also maps the ADF longitudinal incoherency. 

Since both the ADF and EDX are incoherent imaging modes, we expanded the atomic lensing model previously developed for ADF to EDX, which could also be applicable for EELS with a large collection angle. The model takes the 3D ordering of the atomic column into account by describing the dynamic diffraction as a superposition of the lensing effects of individual atoms focusing the incident electrons. The speed and accuracy of the atomic lensing model were compared against multisclice and PRISM algorithms. We demonstrated that this model can reliably predict EDX values for a Pt@Ag core-shell nanoparticle and for an Al-Ag-Pt-Au high entropy alloy up to 25 atoms (10 nm). Beyond this thickness, the contribution of neighbouring columns becomes significant. This method opens opportunities to quantify atomic resolution EDX and to explore the enormous amount of ordering possibilities of heterogeneous materials with multiple elements.

\section*{Acknowledgement}
The authors acknowledge financial support from the Research Foundation Flanders (FWO, Belgium) through Project No.G.0502.18N and a post-doctoral grant to ADB. This project has received funding from the European Research Council (ERC) under the European Union’s Horizon 2020 research and innovation programme (Grant Agreement No. 770887 PICOMETRICS and No. 823717 ESTEEM3). ZZ acknowledges the consultation with Scott Findlay and Les J. Allen for muSTEM calculations. 
% \bibliographystyle{elsarticle-num}
% \bibliography{references}

\end{document}